# Low-symmetry monoclinic phase stabilized by oxygen octahedra rotations in thin strained $Eu_xSr_{1-x}TiO_3$ films

Anna N. Morozovska[1,2], Yijia Gu[3], Victoria V. Khist[2], Maya D. Glinchuk[2], Long-Qing Chen[3*], Venkatraman Gopalan[3†], and Eugene A. Eliseev[2‡]

[1] Institute of Physics, NAS of Ukraine, 46, pr. Nauki, 03028 Kiev, Ukraine

[2] Institute for Problems of Materials Science, NAS of Ukraine, Krjijanovskogo 3, 03142 Kiev, Ukraine

[2] Department of Materials Science and Engineering, Pennsylvania State University, University Park, Pennsylvania 16802, USA

Using Landau-Ginzburg-Devonshire theory and phase field modeling, we explore the complex interplay between the long-range structural order parameter (oxygen octahedron rotations) and polarization in $Eu_xSr_{1-x}TiO_3$ thin epitaxial films. In biaxially tensile strained films, we discover the presence of a low symmetry monoclinic phase with in-plane ferroelectric polarization that is stabilized by antiferrodistortive oxygen octahedra tilts. The monoclinic phase is stable in a wide temperature range, and is characterized by the large number of energetically equivalent polar and structural twin domains, which stimulates easy twinning of the film and thus enhances its effective piezoelectric response. The flexoelectric coupling and rotostriction give rise to additional spontaneous polarization, piezo- and pyro-electricity in the ferroelastic twin boundaries.

[*] vxg8@psu.edu
[†] lqc3@ems.psu.edu
[‡] eugene.a.eliseev@gmail.com

# I. INTRODUCTION

Strain tuning of epitaxial and commensurate complex oxide thin films on appropriate substrates allows for the emergence of a broad range of new properties, as well as tuning of existing properties [1]. These properties include ferroelectricity [2, 3], magnetism [4], octahedral tilts [5], and multiferroicity [6]. Strained films properties can strongly differ from the bulk material properties; and in particular new phases with strong polar or magnetic long-range order, principally not encountered in the bulk, can be caused by misfit strain in ferroelastics and quantum paraelectrics [1-5, 7].

A classical illustration is thoroughly studied strained $SrTiO_3$ films, which have been shown to possess all intriguing properties, including octahedral tilts and ferroelectricity up to 400 K at misfit strain of about 1% or more [7, 8, 9], superconductivity [10], and surprisingly, magnetism [11], whose origin remains uncertain [12]. Also the ferroelectric and structural domain morphologies were predicted at selected values of misfit strain [9]. At the same time, bulk $SrTiO_3$ is a nonmagnetic quantum paraelectric [13] with antiferrodistortive (AFD) structural order below 105 K [14, 15, 16]. Relatively newly and actively studied strained $EuTiO_3$ films becomes strong ferroelectric ferromagnet under epitaxial strains exceeding 1% [17, 18], in contrast to bulk quantum paraelectric $EuTiO_3$ that is a low temperature antiferromagnet [19, 20]. It also exhibits antiferrodistortion at temperatures below ~281 K [21, 22, 23, 24] and is paraelectric at higher temperatures.

Strained films of quantum paraelectric solid solutions, $Eu_xSr_{1-x}TiO_3$, that combine both these material systems, can be intriguing for fundamental studies of structural and polar modes interactions and possibly for multiferroic applications. It appears that $Eu_xSr_{1-x}TiO_3$ nanoparticles and thin films physical properties are very poorly studied, in contrast to widely studied properties of $SrTiO_3$ films. Moreover, the structural AFD and other physical properties of bulk solid solution $Eu_xSr_{1-x}TiO_3$ have been studied experimentally only in one recent work [25]. Possible multiferroic properties of $Eu_xSr_{1-x}TiO_3$ nanotubes and nanowires [26] were predicted using Landau-Ginzburg-Devonshire (LGD) theory, but the important impact of the structural AFD order parameter was considered in the oversimplified scalar approximation. Since the true AFD order parameter is the axial vector of the oxygen octahedral tilt angles [27] that strongly influences the phase diagrams, polar and pyroelectric properties of quantum paraelectric $SrTiO_3$ [16, 28] its interfaces [29] and thin films [7, 9, 30], the strong influence is expected for $Eu_xSr_{1-x}TiO_3$ films.

Strained films phase diagrams are typically much richer with different phases that may be absent in the bulk material. For instance, monoclinic phases having in-plane and out of plane polarization components with different amplitudes were predicted theoretically in epitaxial $BaTiO_3$ films [31, 32, 33]. For the incipient ferroelectric $SrTiO_3$ only tetragonal and orthorhombic phases was predicted [7, 9]. Phase diagrams of $Eu_xSr_{1-x}TiO_3$ films have not been studied previously and so the possibility of low symmetry monoclinic phase stability in the films can be of particular interest due to the high tunability and versatility of possible ferroelectric and ferroelastic twin walls orientations in the phase. Such highly tunable states result in large enhancements in piezoelectric coefficients observed in the presence of domain and twin walls. Also it was shown theoretically that the flexoelectric coupling combined with a rotostriction effect can lead to a spontaneous polarization within the ferroelastic twin walls [28] and the twin walls – surface junctions [34]. Moreover, recent experimental study [35] of domain wall damping and elastic softening of twin walls in $SrTiO_3$ reveals their polar properties and provides evidence that the ferroelastic domain walls become ferroelectric at low temperatures.

In this work we study the long-range structural and polar ordering as well as the phase diagrams of $Eu_xSr_{1-x}TiO_3$ thin strained films using LGD theory and phase field modeling. We pay special attention to the origin of a low symmetry ferroelectric monoclinic phase, whose stability region rapidly enlarges as the Eu content increase, and to the flexo-roto coupling that appears at the twin walls. This paper is organized as follows. The LGD potential for $Eu_xSr_{1-x}TiO_3$ is listed in Section II. Phase diagrams, structural and polar properties of $Eu_xSr_{1-x}TiO_3$ thin films are presented and analyzed in the Section III. Results are summarized in the Section IV. Material parameters of $Eu_xSr_{1-x}TiO_3$ and calculation details are given in the Supplementary Materials.

## II. 2-4 LANDAU-GINZBURG-DEVONSHIRE POTENTIAL FOR $Eu_xSr_{1-x}TiO_3$

Let us consider a short-circuited $Eu_xSr_{1-x}TiO_3$ film of thickness $h$ that is clamped on a rigid substrate (**Fig.1**). Mismatch strain $u_m$ can exist at the interface. AFD structural order can be characterized by spontaneous displacement of oxygen atoms, which could be viewed as oxygen octahedron rotation (measured as displacement of oxygen ion or "tilts"), described by an axial vector $\Phi_i$ ($i$=1, 2, 3) [27]. Polarization vector is $P_i$. To concentrate the readers' attention of the polar-structural subsystem only, below we regard the temperatures more than 50 K, for which any magnetism and magnetoelectricity originating from Eu ions are absent. Low temperature magnetic properties of $EuTiO_3$ were studied in Refs.[17, 18, 36].

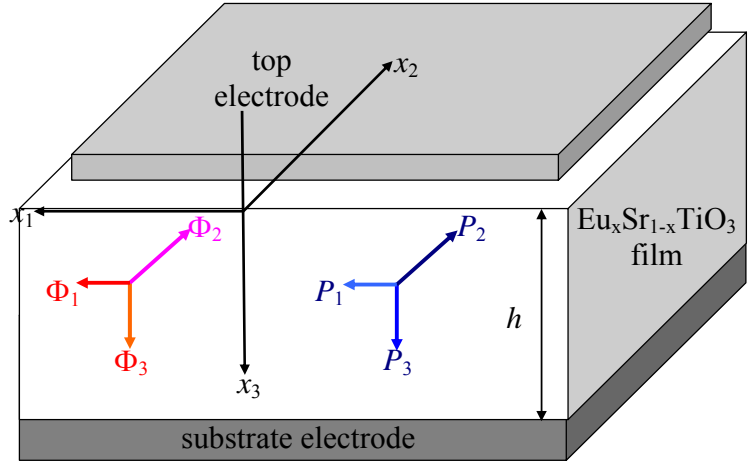

**Figure 1.** Schematics of a short-circuited $Eu_xSr_{1-x}TiO_3$ film clamped on a rigid substrate. Epitaxial misfit strain $u_m$ can exist at the film/substrate interface. Are

2-4-power LGD Gibbs potential density of $Eu_xSr_{1-x}TiO_3$ solid solution depends on the polarization and oxygen octahedra tilt vectors as:

$$G = G_S + \int_0^h \left( G_{grad} + G_{flexo} + G_{elastic} + G_{P\Phi} \right) dx_3 \qquad (1)$$

Where $G_S = a_i^S \left( P_i^2(0) + P_i^2(h) \right) + b_i^S \left( \Phi_i^2(0) + \Phi_i^2(h) \right)$ is the surface term; $G_{grad} = \dfrac{g_{ijkl}}{2} \left( \dfrac{\partial P_i}{\partial x_j} \dfrac{\partial P_k}{\partial x_l} \right) + \dfrac{v_{ijkl}}{2} \left( \dfrac{\partial \Phi_i}{\partial x_j} \dfrac{\partial \Phi_k}{\partial x_l} \right)$ is the gradient energy, $G_{flexo} = \dfrac{F_{ijkl}}{2} \left( \sigma_{ij} \dfrac{\partial P_k}{\partial x_l} - P_k \dfrac{\partial \sigma_{ij}}{\partial x_l} \right)$ is the flexoelectric energy, $F_{ijkl}$ is the forth-rank tensor of flexoelectric coupling experimentally determined for $SrTiO_3$ by Zubko et al in a wide temperature range [37]. $G_{elastic}$ is elastic energy, and $G_{P\Phi}$ is polarization-and-tilt-dependent term. The form of $G_{grad} + G_S$ is the same as listed in Ref.[36]. The elastic energy is given as $G_{elastic} = -s_{ijkl}\sigma_{ij}\sigma_{kl}/2$, where elastic compliances $s_{ijkl}(x) = x s_{ijkl}^{EuTiO_3} + (1-x) s_{ijkl}^{SrTiO_3}$; $\sigma_{ij}$ is the elastic stress tensor. The polarization and structural parts of the LGD-potential density for cubic m3m parent phase is [26]:

$$G_{P\Phi} = \begin{pmatrix} \alpha_P P_i^2 + \beta_{Pij} P_i^2 P_j^2 - Q_{ijkl}\sigma_{ij} P_k P_l + \alpha_\Phi \Phi_i^2 \\ + \beta_{\Phi ij} \Phi_i^2 \Phi_j^2 - R_{ijkl}\sigma_{ij}\Phi_k \Phi_l + \dfrac{\xi_{ik}}{2}\Phi_i^2 P_k^2 \end{pmatrix} \qquad (2)$$

The biquadratic coupling between the structural order parameter $\Phi_i$ and polarization components $P_i$ are defined by the tensor $\xi_{ik}$. [16, 38]. The biquadratic coupling term was later regarded as Houchmandazeh-Laizerowicz-Salje (HLS) coupling [39]. The coupling was considered as the reason of magnetization appearance inside the ferromagnetic domain wall in non-ferromagnetic media [40]. Biquadratic coupling tensor and higher order expansion coefficients are regarded composition dependent: $\beta_{P,\Phi}(x) = x\beta_{P,\Phi}^{EuTiO_3} + (1-x)\beta_{P,\Phi}^{SrTiO_3}$, $\xi_{ij}(x) = x\xi_{ij}^{EuTiO_3} + (1-x)\xi_{ij}^{SrTiO_3}$. $Q_{ijkl}(x) = xQ_{ijkl}^{EuTiO_3} + (1-x)Q_{ijkl}^{SrTiO_3}$ and $R_{ijkl}(x) = xR_{ijkl}^{EuTiO_3} + (1-x)R_{ijkl}^{SrTiO_3}$ are the electrostriction and rotostriction tensors components respectively, which also depend linearly on the composition $x$. Coefficient $\alpha_P(T,x)$ depend on temperature in accordance with Barrett law [41] and composition $x$ of Eu$_x$Sr$_{1-x}$TiO$_3$ solid solution as

$$\alpha_P(T,x) = x\alpha_P^{EuTiO_3}(T) + (1-x)\alpha_P^{SrTiO_3}(T) \qquad (3a)$$

$$\alpha_P(T) = \alpha_T^{(P)}\left(T_q^{(P)}/2\right)\left(\coth\left(T_q^{(P)}/2T\right) - \coth\left(T_q^{(P)}/2T_c^{(P)}\right)\right). \qquad (3b)$$

Temperatures $T_q^{(P)}$ are so called quantum vibration temperatures for SrTiO$_3$ and EuTiO$_3$ respectively, related with either polar modes, $T_c^{(P)}$ are the "effective" Curie temperatures corresponding to polar soft modes in bulk EuTiO$_3$ and SrTiO$_3$.

Note, that recently Zurab Guguchia et al. [25] experimentally observed a nonlinear composition dependence of temperature of transition from cubic non-AFD and tetragonal AFD phase included in Eq.(3b) as $T_S(x) \approx 113.33 + 390.84x - 621.21x^2 + 398.87x^3$ [42]. To account for the experiment and Barrett law coefficient $\alpha_\Phi(T,x)$ depend on temperature and composition $x$ of Eu$_x$Sr$_{1-x}$TiO$_3$ solid solution as

$$\alpha_\Phi(T,x) = \alpha_T^{(\Phi)}(x)\left(T_q^{(\Phi)}(x)/2\right)\left(\coth\left(T_q^{(\Phi)}(x)/2T\right) - \coth\left(T_q^{(\Phi)}(x)/2T_S(x)\right)\right) \qquad (4)$$

For other parameters in Eq.(4) we used a sort of linear extrapolation: $\alpha_T^{(\Phi)}(x) = x \cdot \alpha_{T\Phi}^{EuTiO_3} + (1-x)\alpha_{T\Phi}^{SrTiO_3}$ and $T_q^{(\Phi)}(x) = x \cdot T_{q\Phi}^{EuTiO_3} + (1-x)T_{q\Phi}^{SrTiO_3}$.

Gibbs potential coefficients become renormalized by the surface tension [36, 26], misfit strains [31] and biquadratic coupling between the structural and polar order parameters [28, 36-30]. LGD expansion $\alpha_{Pi}^*$, $\beta_{Pij}^*$, $\alpha_{\Phi i}^*$, $\beta_{\Phi ij}^*$ and coupling coefficients $\xi_{ij}^*$ renormalized by misfit strain via electrostriction and rotostriction mechanism, and extrapolation lengths via finite size and gradient effects are listed in **Appendix A, Suppl. Mat.**. To account for misfit dislocations appearance at film thickness more than $h_d$, effective misfit strain [43] can be introduced as $u_m^*(h) = u_m$ at $\frac{h_d}{h} < 1$ and $u_m^*(h) = u_m\frac{h_d}{h}$ at $\frac{h_d}{h} \geq 1$. In the numerical calculations, which results are presented below, we regard that extrapolation lengths are much greater than the film thickness $h$.

### III. PHASE DIAGRAMS OF Eu$_x$Sr$_{1-x}$TiO$_3$ THIN FILMS: NOVEL PHASES

Numerical simulations of the Eu$_x$Sr$_{1-x}$TiO$_3$ bulk and thin films polar, structural properties and phase diagrams were performed as a function of temperature $T$, composition $x$ and misfit strain $u_m^*$. The numerical values of material parameters included into 2-4-LGD-expansion are listed in the **Table S1, Suppl. Mat**.

$Eu_xSr_{1-x}TiO_3$ bulk and thin films phase diagrams are presented in the **Figures 2-3.** Phase designations indicate all polar ($P_i$) and structural ($\Phi_i$) order parameters components which are nonzero in the phase. For instance designation $P_3\Phi_3$ corresponds to the tetragonal phase with $P_3 \neq 0$ and $\Phi_3 \neq 0$. Abbreviation "ortho" stands for the orthorhombic phase with $P_1 = P_2 \neq 0$ and $\Phi_1 = \Phi_2 \neq 0$. Abbreviation "mono" stands for the low symmetry monoclinic phase with $P_1 \neq P_2 \neq 0$ and $\Phi_1 = \Phi_2 \neq 0$. Abbreviation "para" stands for the paraelectric non-structural phase. Boundary between AFD phases $\Phi_1$ and $\Phi_3$ is shown by a thick dashed line.

$Eu_xSr_{1-x}TiO_3$ bulk, unstrained ($u_m$=0), weakly strained ($|u_m| \leq 0.01\%$) and strongly strained ($|u_m| = 2\%$) thin films phase diagrams in coordinates temperature-composition are shown in the **Figures 2a, 2b** and **2c-d** correspondingly. Two intriguing features were calculated, namely a morphotropic-like boundary between AFD in-plane and out-of-plane phases and a thermodynamically stable monoclinic phase.

The boundary between AFD phases $\Phi_1$ and $\Phi_3$, that exists in the weakly strained films only, i.e. at $|u_m| \leq 0.01\%$, is morphotropic-like and becomes spontaneously twinned. Note that the phases $\Phi_1$ and $\Phi_3$ are indistinguishable in the bulk, since they have the same energy and represent the different variants of the same bulk tetragonal phase. However, internal biaxial stresses exist in the thin epitaxial films clamped to a rigid substrate even at zero misfit strains (see **Appendix A, Suppl. Mat**). The stresses leads to the renormalization of the coefficients $\beta^*_{Pij}$ and $\beta^*_{\Phi ij}$ even at $u_m = 0$, meanwhile $\alpha^*_{\Phi i} = \alpha_{\Phi i}$ at $u_m = 0$ in Eq.(2). Thus the clamping-related stresses break the symmetry between the in-plane and the out-of-plane direction making the AFD phases with the order parameter pointed along these two directions thermodynamically non-equivalent.

Analytical expressions for the order parameters in the monoclinic phase with $P_1 \neq P_2 \neq 0$ and $\Phi_1 \neq \Phi_2 \neq 0$ are listed in **Appendix B, Suppl. Mat.** The degree of "monoclinity" can be estimated as $P_1^2 - P_2^2 = a_m\sqrt{\dfrac{P_m^4 - \phi^2 \Phi_m^4}{a_m^2 - \phi^2}}$ and $\Phi_1^2 - \Phi_2^2 = \sqrt{\dfrac{P_m^4 - \phi^2 \Phi_m^4}{a_m^2 - \phi^2}}$, where $P_m \equiv \sqrt{P_1^2 + P_2^2}$ and $\Phi_m \equiv \sqrt{\Phi_1^2 + \Phi_2^2}$; the cumbersome expressions for $a_m$, $\phi^2$, $P_m^4$ are given in **Appendix B.** The monoclinic phase appeared at very small x. A rather thin monoclinic region appears in pure $SrTiO_3$ (x=0) as shown in **Figure C1, Appendix C, Suppl. Mat.** The monoclinic region strongly increases with Eu content x increase and temperature decrease. One of the necessary conditions of the monoclinic phase appearance is simultaneous presence of spontaneous polarization and tilt is the *negative sign* of biquadratic coupling tensor coefficients $\xi_{ik}$ (see **Table 1**). Also LGD-expansion coefficients $\alpha^*_{Pi}$ and $\alpha^*_{\Phi i}$ should be negative, but these conditions could be readily reached in the strained films, since the former coefficients could be essentially renormalized by misfit strains. The conditions $\xi^*_{ij} < 0$ are valid if $\xi_{ij} < 0$, since the renormalization of $\xi_{ik}$ by misfit effect is usually small. According to our calculations, exactly opposite signs for the coupling $\xi_{ik}$ in SrTiO$_3$ versus EuTiO$_3$ can explain the monoclinic phase region increasing under the increase of Eu content, *x*. Thus we can conclude that simultaneous presence of both octahedra tilts and polarization in epitaxial Eu$_x$Sr$_{1-x}$TiO$_3$ films stabilize in-plane monoclinic phase at moderate and high tensile strains $u_m > 1\%$.

**Table 1**

| AFD material | Biquadratic coupling type with respect to the monoclinic phase origin | Refs. |
|---|---|---|
| SrTiO$_3$ | unfavourable | e.g. [7] |
| EuTiO$_3$ | favourable | Our fit |
| CaTiO$_3$ | unfavourable | Yijia Gu et al. [5] |
| PbZr$_x$Ti$_{1-x}$O$_3$ | favourable | Haun et al. [38] |

**Figures 3** show phase diagrams of Eu$_x$Sr$_{1-x}$TiO$_3$ thin films in coordinates misfit strain–composition (plots **a,b**) and temperature - misfit strain (plots **c,d**). In the coordinates misfit strain–composition the boundary $\Phi_1/\Phi_3$ corresponds to very small misfit strains $|u_m| \leq 0.01\%$ and is almost independent on the composition, until the transition from the AFD to para-phase occurs. The para-phase region increase with the temperature increase (compare **Figs. 3a** and **3b**). Compressive strains stabilize out-of-plane tetragonal and orthorhombic polar and structural phases as anticipated from earlier studies of SrTiO$_3$ [7]. Pertsev et al. [7] calculated the temperature - misfit strain phase diagram of homogeneous epitaxial films of SrTiO$_3$. They predicted the existence of both pure AFD phases and AFD-FE phases as well as the anomalies of the internal misfit stresses at the transition points, but did not report any monoclinic phase. In accordance with our calculations, an ultra-thin monoclinic region of the low symmetry monoclinic phase appears at tensile strains more than 1% for pure SrTiO$_3$ only appears (see **Figure B1 in Appendix B, Suppl. Mat.).** The monoclinic phase region strongly increases with Eu content x increase and temperature decrease. Different orthorhombic phases ( $P_1 = P_2 \neq 0$ and $\Phi_1 = \Phi_2 \neq 0$, and $P_1 = P_2 \neq 0$ ) dominate for small *x*. With *x* increase the monoclinic phase fills the orthorhombic phase region.

The monoclinic phase exists in tensile strained EuTiO$_3$ films ($u_m \approx 2\%$) up to the temperatures 400 K and higher.

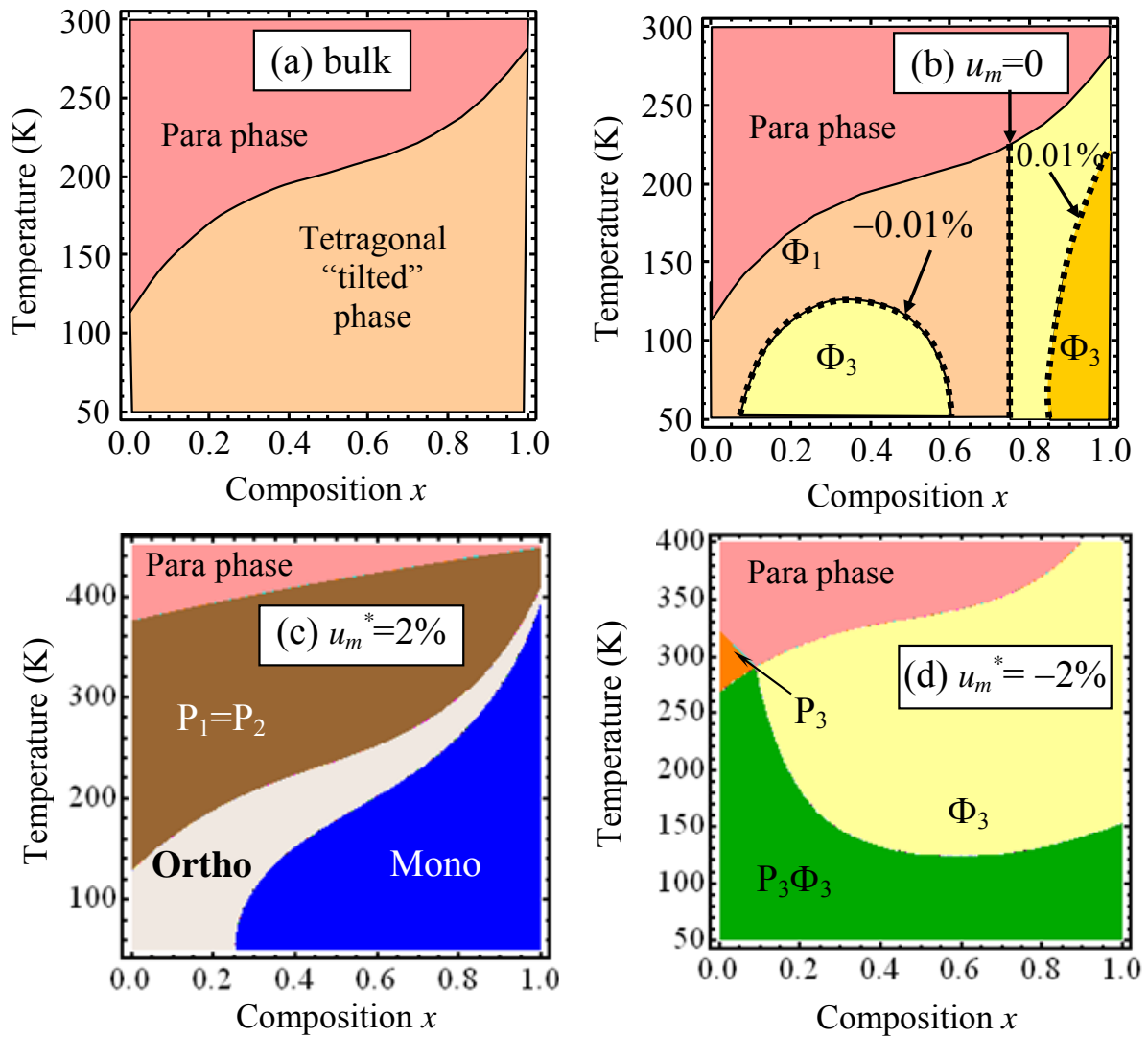

**Figure 2.** Temperature - composition phase diagrams of Eu$_x$Sr$_{1-x}$TiO$_3$ calculated for **(a)** bulk and **(b)** thin film on the matched substrate corresponding to zero misfit $u_m$=0 (vertical boundary $\Phi_1/\Phi_3$ ), $u_m^*$= –0.01% (left $\Phi_3$ region), $u_m^*$= +0.01% (right $\Phi_3$ region), other substrates corresponding to misfits $u_m^*$= +2% **(c)**, $u_m^*$=–2% **(d).**

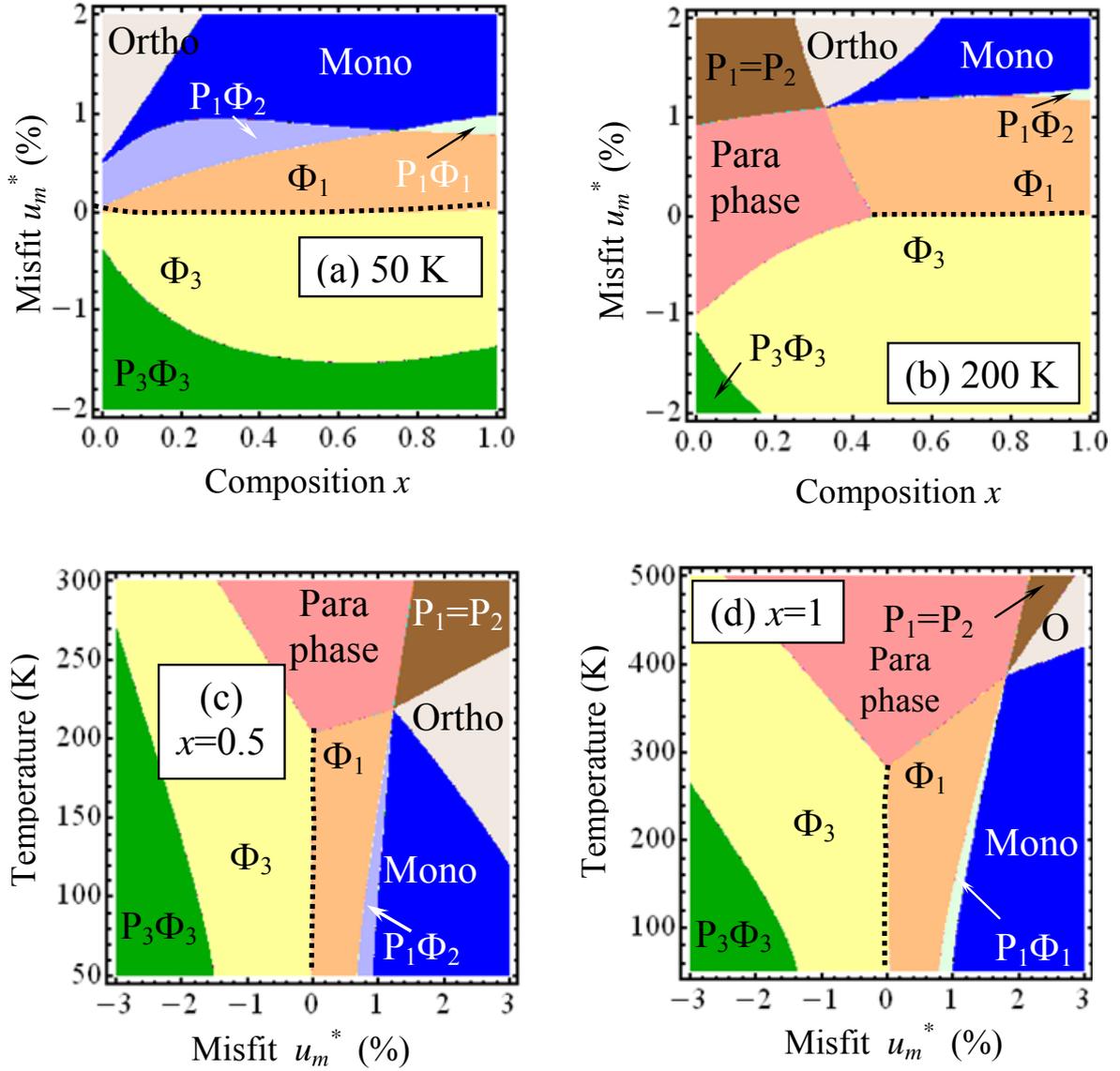

**Figure 3.** The misfit strain–composition phase diagrams of $Eu_xSr_{1-x}TiO_3$ thin films at temperature 50 K **(a)** and 200 K **(b)**. Temperature - misfit strain phase diagrams of $Eu_xSr_{1-x}TiO_3$ thin films for composition $x=0.5$ **(c)** and $x=1$ **(d)**.

One can see from the **Figures 4a,b** that the linear dielectric permittivity demonstrates typical peculiarities (jump or divergences) in the phase transition points. Nonzero component of permittivity tensor $\varepsilon_{12}$ and the condition $\varepsilon_{11} \neq \varepsilon_{22} \neq \varepsilon_{33}$ are the unique features of the monoclinic phase realization in the tensiled $Eu_xSr_{1-x}TiO_3$ films. The dielectric anisotropy factors $\varepsilon_{11}/\varepsilon_{22}$ and $\varepsilon_{22}/\varepsilon_{33}$ is of several to several hundreds of times for the e.g. $Eu_{0.5}Sr_{0.5}TiO_3$ films with effective misfit strain $u_m^* = +1\%$. So the monoclinicity degree strongly affects the dielectric permittivity anisotropy. Temperature dependence of piezoelectric constants in the monoclinic phase ($0 < T < 380$ K) of tensile strained $EuTiO_3$ film is shown in **Figure 4c.** The values characterize the piezoelectric response components of a single-domain film. Some components of piezoelectric response reveal drastic enhancement in the vicinity of the twin walls (**Figure 4d**) promising the possible appearance of new highly tunable stable states in incipient ferroelectrics.

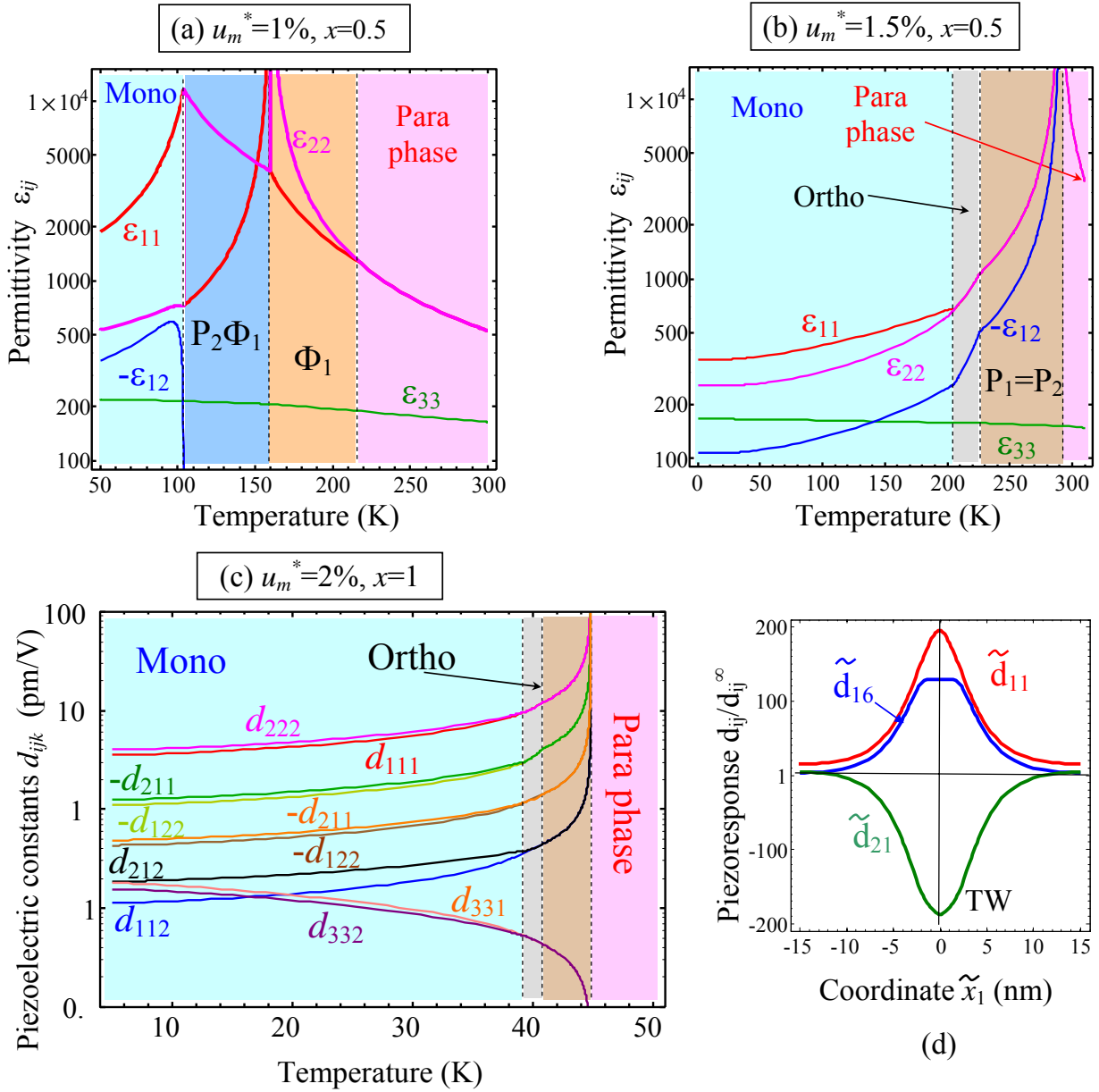

**Figure 4.** Temperature dependence of dielectric permittivity of $Eu_{0.5}Sr_{0.5}TiO_3$ films calculated for tensile misfit strains $u_m^* = +1\%$ (**a**), $u_m^* = +1.5\%$ (**b**). Temperature dependence of piezoelectric constants of tensiled $EuTiO_3$ film, $u_m^* = +2\%$ (**c**). Different phases are separated by the vertical lines. Phase designations are the same as in **Figure 3**. Schematic profiles of several piezoelectric response components across the twin walls (**d**).

Note that the monoclinic phases are possible in the non-structural ferroelectric films only when the corresponding LGD expansion includes the terms of order higher than 6$^{th}$ powers on polarization [44], so that monoclinic phase could appear as the intermediate phase between the higher order phases [45]. In fact, the full set of LGD expansion coefficients with the order up $8^{th}$ powers on polarization was determined only for BaTiO$_3$ [46]; nevertheless the reliability of this set is disputed [47]. Monoclinic phase was found in Pb(Zr,Ti)O$_3$ by Noheda et al. [48] at the morphotropic boundary. It was demonstrated [49] that the monoclinic phase is accompanied by the octahedral tilts, at least at lower temperatures. Monoclinic phase was predicted in the superlattice of BaTiO$_3$ and SrTiO$_3$ films [50], however it was a consequence of complex domain structure in the multilayered ferroelectric film. In the case of Eu$_x$Sr$_{1-x}$TiO$_3$ we start from 2-4-LGD expansion for polarization, but the similar 2-4-LGD expansion for the structural order parameter increase the effective power of the coupled Euler-Lagrange equations up the $8^{th}$ power.

The monoclinic phase is characterized by the high number of energetically equivalent tilt and polarization domains, leading to the origin of not less than 16 elastic and 8 ferroelectric different twin types. Naturally, it is 2 times higher than for the orthorhombic phase with $P_1 = P_2 \neq 0$ and $\Phi_1 = \Phi_2 \neq 0$. High number of possible domains pairs (mostly twins) strongly favours both the spontaneous twinning of the strained film and higher piezoelectric response, i.e. enhances the film electromechanical tunability.

The film becomes spontaneously twinned in its vicinity of the morphotropic boundaries $\Phi_1/\Phi_3$ as well as in the monoclinic phase with $P_1 \neq P_2 \neq 0$ and $\Phi_1 \neq \Phi_2 \neq 0$. Immediately, every elastic twin boundary (TB) gives rise to the spontaneous polarization, piezo- and pyro-electricity due to the presence of rotostriction and flexoelectric coupling [28]. Recent experiment [35] supports the theoretical assumption. Actually Scott et al [35] studied the damping and elastic softening of twin walls in bulk SrTiO$_3$ and showed that ferroelastic domain walls become ferroelectric at low temperatures.

The joint action of rotostriction and flexoelectric coupling [28] causes the inhomogeneous strain $u_{ij}(\mathbf{r}) \propto R_{mnpq} \partial(\Phi_p \Phi_q)/\partial x_l$ across the structural TB, that induces the polarization variation $\delta P_i(\mathbf{r})$, piezoelectric response and pyroelectric response increase, $\delta d_{ijk}(\mathbf{r})$ and $\delta \Pi_i(\mathbf{r})$ due to the direct flexoelectric effect:

$$\delta P_i(\mathbf{r}) \propto -\alpha_{iv}^{-1} f_{mnvl} R_{mnpq} \frac{\partial(\Phi_p \Phi_q)}{\partial x_l}, \qquad (5a)$$

$$\delta d_{ijk}(\mathbf{r}) \approx 2\varepsilon_0 \varepsilon_{im} Q_{mkjl} \delta P_l(\mathbf{r}), \qquad \delta \Pi_i(\mathbf{r}) = -\gamma_{ij} \frac{\partial}{\partial T} \delta P_j(\mathbf{r}). \qquad (5b)$$

The term $f_{mnvl}$ denotes direct flexoelectric tensor, $f_{ijkl} = c_{ijmn} F_{mnkl}$; $\gamma_{ij}$ is the pyroelectric coefficients tensor; $\varepsilon_0 = 8.85 \times 10^{-12}$ F/m is the universal dielectric constant, $\varepsilon_{ij}$ is the relative dielectric permittivity. Note, that Eq.(5a) is valid only for zero electric field, including both external and depolarization fields. The estimations give $0.5 - 5$ C/m$^2$ for $\delta P_i(\mathbf{r})$, 10 pm/V for $\delta d_{ijk}(\mathbf{r})$ and $(5-50)10^{-6}$ C/m$^2$K for $\delta\Pi_i(\mathbf{r})$ depending on temperature and content x. The numerical values are in agreement with previous studies of roto-flexo effect impact [28, 29, 30]. Note, that there are other possible reasons for polar surface states in nonpolar materials such as SrTiO$_3$, primary such as surface piezoelectricity [51, 52, 53, 54]. In numbers, Dai et al [54] obtained the surface polarization ~(0.07 - 0.02) μC/cm$^2$ for SrTiO$_3$ entire the temperature range. Roto-flexo effect can lead to higher values ~(0.5-5) μC/cm$^2$, but it exists in the AFD phases only.

The spontaneous polarization induced by the tilt gradient in the vicinity of SrTiO$_3$ TB was calculated by the phase field modelling. Profiles of the tilts $\widetilde{\Phi}_1$ and $\widetilde{\Phi}_2$ and polarization $\widetilde{P}_1$ and $\widetilde{P}_2$ components across the easy and hard TB calculated in the rotated frame $\{\widetilde{x}_1, \widetilde{x}_2\}$ are shown in the **Figure 5**. For **hard TB**, shown in the **Figure 5c-d**, $\widetilde{P}_1$ is odd and $\widetilde{P}_2$ is even. The even Bloch-type component $\widetilde{P}_2$ flips when $\widetilde{\Phi}_1$ flips, as one can conclude comparing the plots (c) and (d). The magnitude of $\widetilde{P}_1$ and $\widetilde{P}_2$ are quite different, this is similar to the hard antiphase boundaries [28]. For **easy TB,** shown in the **Figure 5e-f**, $\widetilde{P}_1$ is odd and $\widetilde{P}_2$ is even. The even $\widetilde{P}_2$ flips when $\widetilde{\Phi}_2$ flips, as one can see comparing the plots (e) and (f). The magnitudes of $\widetilde{P}_1$ and $\widetilde{P}_2$ are similar. These results are in a semi-quantitative agreement with previous analytical results [28], however one interesting parity-related effect that was not reported previously is evident. It is the flip of the even Bloch-type polarization distribution $\widetilde{P}_2$ with the change of the Ising-type tilt component $\widetilde{\Phi}_1$ sign. At that the odd component of polarization profile is independent on the tilt sign. The physical explanation of the effect can be given after the analyses of the symmetry of the inhomogeneous strain $\widetilde{u}_{ij}(\mathbf{r}) \propto \widetilde{R}_{mnpq} \partial(\widetilde{\Phi}_p \widetilde{\Phi}_q)/\partial\widetilde{x}_l$ that is the seeding for the components $\widetilde{P}_1$ and $\widetilde{P}_2$ in the vicinity of TB. The seeding in the Euler-Lagrange equations [28] for polarization components are $\widetilde{P}_1 \propto \partial(\widetilde{\Phi}_1^2)/\partial\widetilde{x}_1$ and $\widetilde{P}_2 \propto \partial(\widetilde{\Phi}_1 \widetilde{\Phi}_2)/\partial\widetilde{x}_1$. So that $\widetilde{P}_1$ is the even function of $\widetilde{\Phi}_1$ and $\widetilde{P}_2$ is the odd function of $\widetilde{\Phi}_1$.

Note, that there were misprints in the gradient coefficients in Ref.[5]. Corrected set of SrTiO$_3$ coefficients are given in the **Table S1, Suppl. Mat.** [55]. Using the correct set of parameters we calculated that the two kinds of twin boundaries have similar wall width (compare left and right column in the **Figure 5**).

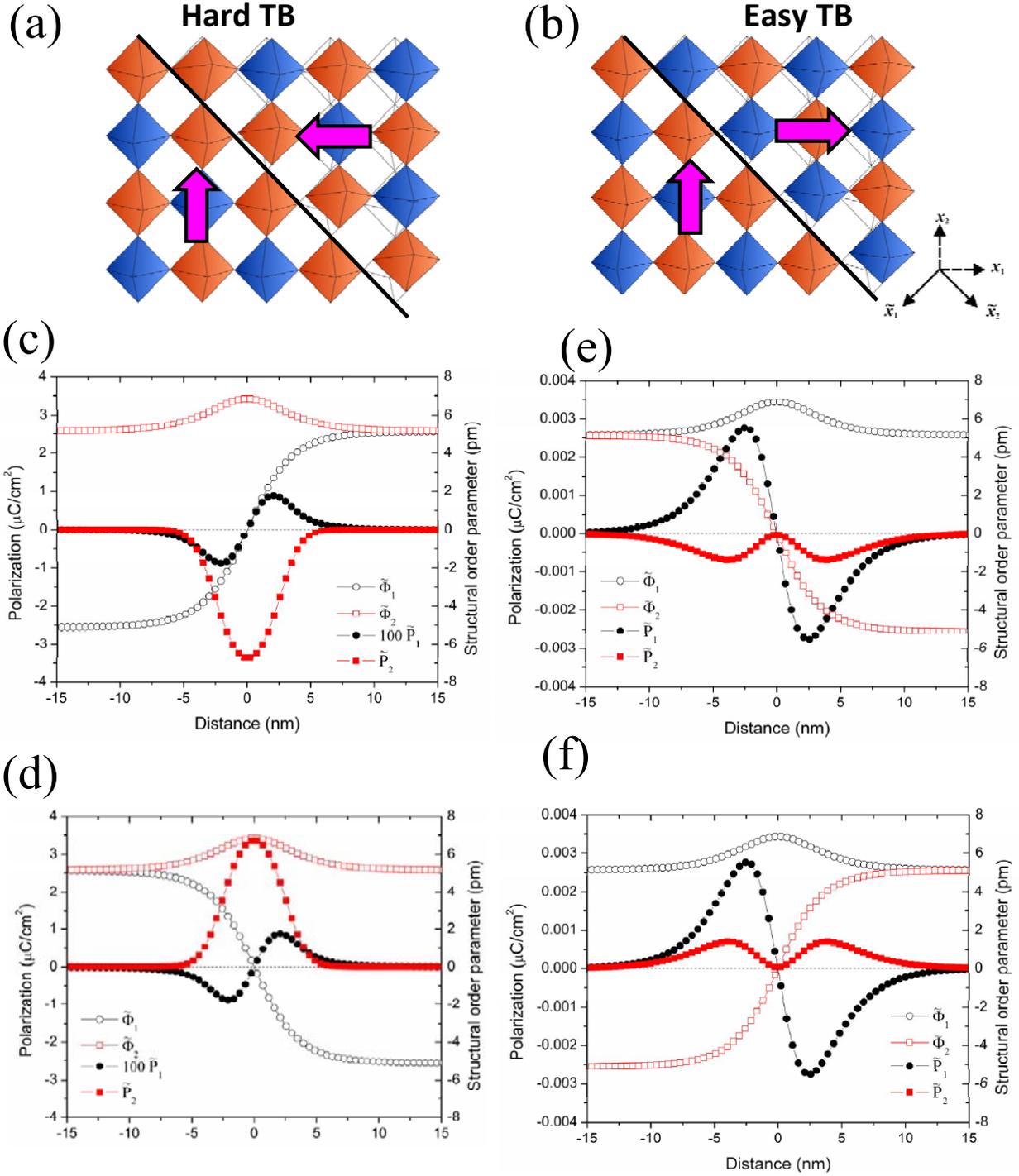

**Figure 5.** (a) Configuration of oxygen octahedrons tilt across the "head-to-head" hard TB. (b) Configuration of oxygen octahedrons tilt across the "head-to-tail" easy TB. (c-f) The spontaneous polarization induced by the tilt gradient in the vicinity of the TB. Results are calculated by the phase-field modelling.

## IV. SUMMARY

Using Landau-Ginzburg-Devonshire 2-4-power expansion and phase field modelling, we study the interplay between the long-range structural order parameter and polarization in $Eu_xSr_{1-x}TiO_3$ thin epitaxial films.

We discover a new low symmetry ferroelectric monoclinic phase that becomes stable in $Eu_xSr_{1-x}TiO_3$ thin epitaxial films at moderate tensile strains. We derived analytical expressions for the spontaneous tilt and ferroelectric polarization vectors in the monoclinic phase and proved that the presence of antiferrodistortive octahedra tilts stabilize the monoclinic phase with in-plane ferroelectric polarization. The monoclinic phase stability region rapidly enlarges with Eu content increase. The phase is thermodynamically stable in a wide temperature range. The monoclinic phase is characterized by a large number of energetically equivalent orientations of the polar and structural order parameters, that enhances its effective piezoelectric response. Since the local elastic field gradients of adjacent domain walls will interact, the appearance of highly tunable piezoelectricity in the incipient ferroelectric films is possible, while it is not expected in bulk $Eu_xSr_{1-x}TiO_3$.

Using phase-field modelling we demonstrate that the flexoelectric coupling and rotostriction give rise to the spontaneous polarization of the elastic twin boundaries due to the intrinsic strain gradient. Phase field results are in a reasonable agreement with analytical results [28]; however one interesting parity-related effect, namely the flip of the Bloch-type polarization distribution with the change of the Ising-type tilt component sign, is seen, that was not reported previously.

**Acknowledgements.** A.N.M. M.D.G., V.V.K and E.A.E. acknowledge State Fund of Fundamental State Fund of Fundamental Research of Ukraine, SFFR-NSF project UU48/002. Y.G, V.G. and L.Q.C. acknowledge NSF-DMR- 0908718, NSF-DMR-0820404, and NSF-DMR-1210588 funds.

# SUPPLEMENTAL MATERIALS
# APPENDIX A. ELASTIC FIELD AND LGD-POTENTIAL WITH RENORMALIZED COEFFICIENTS

Neglecting gradient and flexoelectric effects elastic fields in the epitaxial thin film clamped on a rigid substrate have the view:

$$\sigma_{33} = 0, \quad \sigma_{13} = 0, \quad \sigma_{23} = 0, \quad \sigma_{12} = -\frac{Q_{44}}{s_{44}} P_1 P_2 - \frac{R_{44}}{s_{44}} \Phi_1 \Phi_2, \quad (A.1a)$$

$$\sigma_{11} = \frac{u_m - Q_{12} P_3^2 - R_{12} \Phi_3^2}{s_{11} + s_{12}} - \frac{1}{2}\left(\frac{Q_{11} + Q_{12}}{s_{11} + s_{12}} - \frac{Q_{11} - Q_{12}}{s_{11} - s_{12}}\right) P_2^2 - \frac{1}{2}\left(\frac{Q_{11} + Q_{12}}{s_{11} + s_{12}} + \frac{Q_{11} - Q_{12}}{s_{11} - s_{12}}\right) P_1^2$$

$$-\frac{1}{2}\left(\frac{R_{11} + R_{12}}{s_{11} + s_{12}} - \frac{R_{11} - R_{12}}{s_{11} - s_{12}}\right) \Phi_2^2 - \frac{1}{2}\left(\frac{R_{11} + R_{12}}{s_{11} + s_{12}} + \frac{R_{11} - R_{12}}{s_{11} - s_{12}}\right) \Phi_1^2 \quad (A.1b)$$

$$\sigma_{22} = \frac{u_m - Q_{12} P_3^2 - R_{12} \Phi_3^2}{s_{11} + s_{12}} - \frac{1}{2}\left(\frac{Q_{11} + Q_{12}}{s_{11} + s_{12}} + \frac{Q_{11} - Q_{12}}{s_{11} - s_{12}}\right) P_2^2 - \frac{1}{2}\left(\frac{Q_{11} + Q_{12}}{s_{11} + s_{12}} - \frac{Q_{11} - Q_{12}}{s_{11} - s_{12}}\right) P_1^2$$

$$-\frac{1}{2}\left(\frac{R_{11} + R_{12}}{s_{11} + s_{12}} + \frac{R_{11} - R_{12}}{s_{11} - s_{12}}\right) \Phi_2^2 - \frac{1}{2}\left(\frac{R_{11} + R_{12}}{s_{11} + s_{12}} - \frac{R_{11} - R_{12}}{s_{11} - s_{12}}\right) \Phi_1^2 \quad (A.1c)$$

$$u_{33} = \left(Q_{11} - \frac{2s_{12} Q_{12}}{s_{11} + s_{12}}\right) P_3^2 + \left(Q_{12} - \frac{s_{12}(Q_{11} + Q_{12})}{s_{11} + s_{12}}\right)(P_2^2 + P_1^2)$$

$$+ \left(R_{11} - \frac{2s_{12} R_{12}}{s_{11} + s_{12}}\right) \Phi_3^2 + \left(R_{12} - \frac{s_{12}(R_{11} + R_{12})}{s_{11} + s_{12}}\right)(\Phi_2^2 + \Phi_1^2) + \frac{2s_{12} u_m}{s_{11} + s_{12}}, \quad (A.2a)$$

$$u_{11} = u_{22} = u_m, \quad u_{12} = 0, \quad (A.2a)$$

$$u_{13} = \frac{Q_{44} P_1 P_3 + R_{44} \Phi_1 \Phi_3}{2}, \quad u_{23} = \frac{Q_{44} P_2 P_3 + R_{44} \Phi_2 \Phi_3}{2}. \quad (A.2c)$$

We regard that the bulk material in the parent high temperature phase has m3m symmetry. For a film on substrate Legendre transformation

$$F_R = G + (\sigma_{11} + \sigma_{22}) u_m \quad (A.3)$$

leads to

$$F_R \cong \alpha_{P1}^*(P_1^2 + P_2^2) + \alpha_{P3}^* P_3^2 + \beta_{P33}^* P_3^4 + \beta_{P13}^*(P_1^2 + P_2^2) P_3^2 + \beta_{P12}^* P_1^2 P_2^2 + \beta_{P11}^*(P_1^4 + P_2^4)$$

$$\frac{u_m^2}{s_{11} + s_{12}} + \xi_{11}^*(P_1^2 \Phi_1^2 + P_2^2 \Phi_2^2) + \xi_{33}^* P_3^2 \Phi_3^2 + \xi_{31}^* P_3^2(\Phi_1^2 + \Phi_2^2)$$

$$+ \xi_{12}^*(P_1^2 \Phi_2^2 + P_2^2 \Phi_1^2) + \xi_{13}^*(P_1^2 + P_2^2) \Phi_3^2 + \xi_{44}^* P_1 P_2 \Phi_1 \Phi_2 + \xi_{44}^{*'}(P_1 \Phi_1 + P_2 \Phi_2) P_3 \Phi_3 \quad (A.4)$$

$$\alpha_{\Phi1}^*(\Phi_1^2 + \Phi_2^2) + \alpha_{\Phi3}^* \Phi_3^2 + \beta_{\Phi33}^* \Phi_3^4 + \beta_{\Phi13}^*(\Phi_1^2 + \Phi_2^2) \Phi_3^2 + \beta_{\Phi12}^* \Phi_1^2 \Phi_2^2 + \beta_{\Phi11}^*(\Phi_1^4 + \Phi_2^4)$$

Glazer symmetry is $a^0 a^0 c-$. For considered geometry, the renormalized coefficients are:

$$\alpha_{P1} = \alpha_P - \frac{Q_{11} + Q_{12}}{s_{11} + s_{12}} u_m^*(h) + \frac{2\pi^2}{\pi^2 h \lambda_{P1} + 2h^2}\left(g_{44} - \frac{f_{44}^2}{c_{44}}\right), \quad (A.5a)$$

$$\alpha_{P3} = \alpha_P - \frac{2Q_{12}}{s_{11} + s_{12}} u_m^*(h) + \frac{2}{(\lambda_{P3} + \sqrt{\varepsilon_0 g_{11}})h}\left(g_{11} - \frac{f_{11}^2}{c_{11}}\right), \quad (A.5b)$$

$$\alpha_{\Phi 1} = \alpha_\Phi - \frac{R_{11} + R_{12}}{s_{11} + s_{12}} u_m^*(h) + \frac{2\pi^2 v_{44}}{\pi^2 h \lambda_{\Phi 1} + 2h^2}, \tag{A.5c}$$

$$\alpha_{\Phi 3} = \alpha_\Phi - \frac{2 R_{12}}{s_{11} + s_{12}} u_m^*(h) + \frac{2\pi^2 v_{11}}{\pi^2 h \lambda_{\Phi 3} + 2h^2}, \tag{A.5d}$$

Extrapolation lengths for polarization $\lambda_{P3} = g_{11}/a_3^S$ and $\lambda_{P1} = g_{44}/a_1^S$ and tilts $\lambda_{\Phi 3} = v_{11}/b_3^S$ and $\lambda_{\Phi 3} = v_{44}/b_1^S$ are introduced. $f_{ijkl} = -c_{ijmn} F_{mnkl}$ is the forth-rank tensor of flexoelectric coupling. The term $\sqrt{\varepsilon_0 g_{11}}$ originated in Eq.(A.5b) from depolarization effects. Effective misfit strain $u_m^*(h) = u_m$ at $\frac{h_d}{h} < 1$ and $u_m^*(h) = u_m \frac{h_d}{h}$ at $\frac{h_d}{h} \geq 1$. Higher terms are:

$$\beta_{P11}^* = \beta_{P11} + \frac{(Q_{11}^2 + Q_{12}^2)s_{11} - 2Q_{11}Q_{12}s_{12}}{2(s_{11}^2 - s_{12}^2)}, \quad \beta_{P33}^* = \beta_{P11} + \frac{Q_{12}^2}{s_{11} + s_{12}}, \tag{A.6a}$$

$$\beta_{P12}^* = \beta_{P12} - \frac{(Q_{11}^2 + Q_{12}^2)s_{12} - 2Q_{11}Q_{12}s_{11}}{s_{11}^2 - s_{12}^2} + \frac{Q_{44}^2}{2s_{44}}, \quad \beta_{P13}^* = \beta_{P12} + \frac{Q_{12}(Q_{11} + Q_{12})}{s_{11} + s_{12}}, \tag{A.6b}$$

$$\beta_{\Phi 11}^* = \beta_{\Phi 11} + \frac{(R_{11}^2 + R_{12}^2)s_{11} - 2R_{11}R_{12}s_{12}}{2(s_{11}^2 - s_{12}^2)}, \quad \beta_{\Phi 33}^* = \beta_{\Phi 11} + \frac{R_{12}^2}{s_{11} + s_{12}}, \tag{A.6c}$$

$$\beta_{\Phi 12}^* = \beta_{\Phi 12} - \frac{(R_{11}^2 + R_{12}^2)s_{12} - 2R_{11}R_{12}s_{11}}{s_{11}^2 - s_{12}^2} + \frac{R_{44}^2}{2s_{44}}, \quad \beta_{\Phi 13}^* = \beta_{\Phi 12} + \frac{R_{12}(R_{11} + R_{12})}{s_{11} + s_{12}}, \tag{A.6d}$$

$$\xi_{11}^* = \xi_{11} + \frac{(Q_{11} + Q_{12})(R_{11} + R_{12})}{2(s_{11} + s_{12})} + \frac{(Q_{11} - Q_{12})(R_{11} - R_{12})}{2(s_{11} - s_{12})}, \tag{A.7a}$$

$$\xi_{12}^* = \xi_{21}^* = \xi_{12} + \frac{(Q_{11} + Q_{12})(R_{11} + R_{12})}{2(s_{11} + s_{12})} - \frac{(Q_{11} - Q_{12})(R_{11} - R_{12})}{2(s_{11} - s_{12})}, \tag{A.7b}$$

$$\xi_{44}^* = \xi_{44} + \frac{R_{44} Q_{44}}{s_{44}}, \quad \xi_{33}^* = \xi_{11} + \frac{2 R_{12} Q_{12}}{s_{11} + s_{12}}, \tag{A.7c}$$

$$\xi_{13}^* = \xi_{12} + \frac{(Q_{11} + Q_{12}) R_{12}}{s_{11} + s_{12}}, \quad \xi_{31}^* = \xi_{12} + \frac{(R_{11} + R_{12}) Q_{12}}{s_{11} + s_{12}}. \tag{7d}$$

The terms in Eq.(A.7) proportional to $u_m$ originated from mismatch strains. Misfit strain and composition dependence of the ordered phases stability can be obtained from the minimization of Eq.(A.4).

Minimization of the free energy (A.4) with respect to $P_i$ and $\Phi_i$ in the single-domain case leads to the system of six coupled algebraic equations:

$$\begin{aligned}
& 2\alpha_{P1}^* P_1 + 2\beta_{P13}^* P_1 P_3^2 + 2\beta_{P12}^* P_1 P_2^2 + 4\beta_{P11}^* P_1^3 + \\
& + 2\xi_{11}^* P_1 \Phi_1^2 + 2\xi_{12}^* P_1 \Phi_2^2 + \xi_{13}^* 2 P_1 \Phi_3^2 + \xi_{44}^* P_2 \Phi_1 \Phi_2 + \xi_{44}^* \Phi_1 P_3 \Phi_3 = 0
\end{aligned} \tag{A.8a}$$

$$2\alpha^*_{P1}P_2 + 2\beta^*_{P13}P_2P_3^2 + 2\beta^*_{P12}P_2P_1^2 + 4\beta^*_{P11}P_2^3 + \\ + 2\xi^*_{11}P_2\Phi_2^2 + 2\xi^*_{12}P_2\Phi_1^2 + \xi^*_{13}2P_2\Phi_3^2 + \xi^*_{44}P_1\Phi_1\Phi_2 + \xi^*_{44}\Phi_2P_3\Phi_3 = 0 \quad \text{(A.8b)}$$

$$2\alpha^*_{P3}P_3 + 4\beta^*_{P33}P_3^3 + 2\beta^*_{P13}(P_1^2 + P_2^2)P_3 + 2\xi^*_{33}P_3\Phi_3^2 \\ + 2\xi^*_{31}P_3(\Phi_1^2 + \Phi_2^2) + \xi^*_{44}(P_1\Phi_1 + P_2\Phi_2)\Phi_3 = 0 \quad \text{(A.8c)}$$

$$2\alpha^*_{\Phi1}\Phi_1 + 2\beta^*_{\Phi13}\Phi_1\Phi_3^2 + 2\beta^*_{\Phi12}\Phi_1\Phi_2^2 + 4\beta^*_{\Phi11}\Phi_1^3 + \\ 2\xi^*_{11}P_1^2\Phi_1 + \xi^*_{31}P_3^2 2\Phi_1 + \xi^*_{12}2P_2^2\Phi_1 + \xi^*_{44}P_1P_2\Phi_2 + \xi^*_{44}P_1P_3\Phi_3 = 0 \quad \text{(A.8d)}$$

$$2\alpha^*_{\Phi1}\Phi_2 + 2\beta^*_{\Phi13}\Phi_2\Phi_3^2 + 2\beta^*_{\Phi12}\Phi_2\Phi_1^2 + 4\beta^*_{\Phi11}\Phi_2^3 + \\ 2\xi^*_{11}P_2^2\Phi_2 + \xi^*_{31}P_3^2 2\Phi_2 + \xi^*_{12}2P_1^2\Phi_2 + \xi^*_{44}P_1P_2\Phi_1 + \xi^*_{44}P_2P_3\Phi_3 = 0 \quad \text{(A.8e)}$$

$$2\alpha^*_{\Phi3}\Phi_3 + 4\beta^*_{\Phi33}\Phi_3^3 + 2\beta^*_{\Phi13}(\Phi_1^2 + \Phi_2^2)\Phi_3 + 2\xi^*_{33}P_3^2\Phi_3 \\ + 2\xi^*_{13}(P_1^2 + P_2^2)\Phi_3 + \xi^*_{44}(P_1\Phi_1 + P_2\Phi_2)P_3 = 0 \quad \text{(A.8f)}$$

Euler-Lagrange equations which describe the system behaviour in the vicinity of the twin boundaries allowing for the tilt and polarization gradient and flexoelectric coupling are listed in the EPAPS to Ref.[28]. The numerical values of material parameters used in the LGD model are listed in the **Table S1**.

Note that Fennie and Rabe [17] predicted the presence of simultaneous ferromagnetic and ferroelectric phases in (001) EuTiO$_3$ thin films under compressive epitaxial strains exceeding 1.2%. Later on Lee *et al.* [18] demonstrated experimentally that EuTiO$_3$ thin films become ferromagnetic at temperatures lower than 4.24 K and ferroelectric at temperatures lower than 250 K under the application of more than 1% tensile misfit strain. Note that Lee *et al.* also made the first principles calculations of ferroelectricity and ferromagnetism in EuTiO$_3$ films, which include the AFD rotations [18, 56]. Later Li et al. [9] considered the range of possible ferroelectric transition temperatures in strained SrTiO$_3$ films with respect to the variation in the reported properties of bulk.

**Table S1**. List of parameters for polarization and tilt dependent part of the free energy

| Parameter (SI units) | SrTiO$_3$ Value | Source and notes | EuTiO$_3$ Value | Source and notes |
|---|---|---|---|---|
| $\varepsilon_b$ | 43 | [a] | 33 | fitting to [b, c] |
| $\alpha_T^{(P)}$ ($\times 10^6$ m/(F K)) | 0.75 | [d] | 1.95 | fitting to [b, c] |
| $T_c^{(P)}$ (K) | 30 | [d] | −133.5 | fitting to [b, c] |
| $T_q^{(P)}$ (K) | 54 | [d] | 230 | fitting to [b, c] |
| $a_{11}^\sigma$ ($\times 10^9$ m$^5$/(C$^2$F)) | 1.696 | [d] | 1.6 | fitting to [e] |
| $a_{12}^\sigma$ ($\times 10^9$ m$^5$/(C$^2$F)) | 3.918 | [f] | 1.4 | fitting to [e] |
| $Q_{ij}$ (m$^4$/C$^2$) | $Q_{11}$=0.046, $Q_{12}$=−0.014, $Q_{44}$=0.019 | Recalculated from [d] | $Q_{11}$=0.10, $Q_{12}$=−0.015, $Q_{44}$=0.019 | fitting to [g] |
| $\alpha_T^{(\Phi)}$ ($\times 10^{26}$ J/(m$^5$ K)) | 9.1 | [h] | 3.91 | fitting to [i, j] |
| $T_c^{(\Phi)}$ (K) | 105 | [h] | 270 | Averaged of. 220 [j], 275 [k] and 282 [i] |
| $T_q^{(\Phi)}$ (K) | 145 | [h] | 205 | fitting to [j] |
| $\beta_{\Phi 11}$ ($\times 10^{50}$ J/m$^7$) | 1.69 | [h] | 0.436 | fitting to [i, j] |
| $\beta_{\Phi 12}$ ($\times 10^{50}$ J/m$^7$) | 3.88 | [h] | 3.88 | estimation |
| $R_{ij}$ ($\times 10^{18}$ m$^{-2}$) | $R_{11}$=8.7, $R_{12}$=−7.8, $R_{44}$=−18.4 | recalculated from [h] | $R_{11}$=5.46, $R_{12}$=−2.35, $R_{44}$=−18.1 | fitting to [j] |
| $\xi_{11}$ ($\times 10^{29}$ (F m)$^{-1}$) $\xi_{12}$ ($\times 10^{29}$ (F m)$^{-1}$) $\xi_{44}$ ($\times 10^{29}$ (F m)$^{-1}$) | 1.744 0.755, −5.85 | [h] | −2.225 0.85, −5.86 | fitting to [b, c] |
| $s_{ij}$ ($\times 10^{-12}$ m$^3$/J) | $s_{11}$=3.52, $s_{12}$=−0.85, $s_{44}$=7.87 | recalculated from [d] | $s_{11}$=3.65, $s_{12}$=−0.85 | first-principles calculations |
| Tilt gradient $v_{ijkl}$ ($10^{10} \times$ J/m$^3$) | $v_{11}$=0.28, $v_{12}$=−7.34, $v_{44}$=7.11 | From [d] | the same as for SrTiO$_3$ | Estimation |
| $g_{ijkl}$ ($10^{-11} \times$ V·m$^3$/C) | $g_{11}$=$g_{44}$=1, $g_{12}$=0.5 | Estimation | the same as for SrTiO$_3$ | Estimation |
| Flexoelectric tensor $F_{ijkl}$ ($10^{-12} \times$ m$^3$/C) | $F_{11}$=−13.80, $F_{12}$=6.66, $F_{44}$=8.48 | recalculated from. [l] | the same as for SrTiO$_3$ | Estimation |

[a] G. Rupprecht and R.O. Bell, Phys. Rev. **135**, A748 (1964).
[b] T. Katsufuji and H. Takagi, Phys. Rev. **B 64**, 054415 (2001).
[c] V. Goian et al. Eur. Phys. J. B **71**, 429–433 (2009).
[d] A.K. Tagantsev, E. Courtens and L. Arzel, Phys. Rev. B, **64**, 224107 (2001).
[e] C.J. Fennie, & K.M. Rabe, Phys. Rev. Lett. **97**, 267602 (2006).
[f] G. Sheng et al. Appl. Phys. Lett. 96, 232902 (2010).
[h] N. A. Pertsev, A. K. Tagantsev, and N. Setter, Phys. Rev. B **61**, R825 (2000).
[g] J.H. Lee et al. Nature, **466**, 954, (2010).
[i] A. Bussmann-Holder, J. Kohler, R. K. Kremer, and J. M. Law. Phys. Rev. **B 83**, 212102 (2011).
[j] Mattia Allieta et al. Phys. Rev. **B 85**, 184107 (2012).
[k] Zurab Guguchia et al. Phys. Rev. **B 85**, 134113 (2012).
[l] P. Zubko et al. Phys. Rev. Lett. **99**, 167601 (2007).

**APPENDIX B. ORDER PARAMETERS FOR LOW SYMMETRY MONOCLINIC PHASE**

In this phase both in-plane components of polarization and tilt are present, but with different amplitudes:

$$P_{1,2} = \sqrt{\frac{P_m^2}{2} \pm \frac{a_m}{2}\sqrt{\frac{P_m^4 - \phi^2 \Phi_m^4}{a_m^2 - \phi^2}}}, \quad \Phi_{1,2} \equiv \sqrt{\frac{\Phi_m^2}{2} \pm \frac{1}{2}\sqrt{\frac{P_m^4 - \phi^2 \Phi_m^4}{a_m^2 - \phi^2}}} \quad (B.1)$$

Here we introduced absolute values of polarization and tilt

$$P_m \equiv \sqrt{P_1^2 + P_2^2} = \left( \frac{-2\alpha_{P1}^*\left(2\beta_{\Phi 11}^* + \beta_{\Phi 12}^* + \frac{\xi_{44}\phi}{2}\right) + 2\alpha_{\Phi 1}^*\left(\xi_{11}^* + \xi_{12}^*\right)}{\left(2\beta_{P11}^* + \beta_{P12}^* + \frac{\xi_{44}}{2\phi}\right)\left(2\beta_{\Phi 11}^* + \beta_{\Phi 12}^* + \frac{\xi_{44}\phi}{2}\right) - \left(\xi_{11}^* + \xi_{12}^*\right)^2} \right)^{1/2} \quad (B.2a)$$

$$\Phi_m \equiv \sqrt{\Phi_1^2 + \Phi_2^2} = \left( \frac{-2\alpha_{\Phi 1}^*\left(2\beta_{P11}^* + \beta_{P12}^* + \frac{\xi_{44}}{2\phi}\right) + 2\alpha_{P1}^*\left(\xi_{11}^* + \xi_{12}^*\right)}{\left(2\beta_{P11}^* + \beta_{P12}^* + \frac{\xi_{44}}{2\phi}\right)\left(2\beta_{\Phi 11}^* + \beta_{\Phi 12}^* + \frac{\xi_{44}\phi}{2}\right) - \left(\xi_{11}^* + \xi_{12}^*\right)^2} \right)^{1/2} \quad (B.2b)$$

Along with two additional parameters:

$$a_m \equiv \frac{P_1^2 - P_2^2}{\Phi_1^2 - \Phi_2^2} = -\frac{2\beta_{\Phi 11}^* - \beta_{\Phi 12}^* - \frac{\xi_{44}\phi}{2}}{\xi_{11}^* - \xi_{12}^*} \quad (B.3)$$

and

$$\phi \equiv \frac{P_1 P_2}{\Phi_1 \Phi_2} = \frac{(2\beta_{P11}^* - \beta_{P12}^*)(2\beta_{\Phi 11}^* - \beta_{\Phi 12}^*) - (\xi_{11}^* - \xi_{12}^*)^2 + \frac{\xi_{44}^2}{4}}{(2\beta_{P11}^* - \beta_{P12}^*)\xi_{44}} \pm$$

$$\pm \frac{\sqrt{\left((2\beta_{P11}^* - \beta_{P12}^*)(2\beta_{\Phi 11}^* - \beta_{\Phi 12}^*) - (\xi_{11}^* - \xi_{12}^*)^2 + \frac{\xi_{44}^2}{4}\right)^2 - (2\beta_{P11}^* - \beta_{P12}^*)(2\beta_{\Phi 11}^* - \beta_{\Phi 12}^*)\xi_{44}^2}}{(2\beta_{P11}^* - \beta_{P12}^*)\xi_{44}} \quad (B.4)$$

It is seen that actually one has two different phases of similar symmetry for two different values of $\phi$.

# APPENDIX C. PHASE DIAGRAMS OF SrTiO$_3$ THIN FILMS

Results of our numerical simulations of the SrTiO$_3$ thin films phase diagrams in coordinates temperature $T$ and misfit strain $u_m$ are shown in **Figure C1**. We put $h < h_d$ and regard that extrapolation lengths are much greater than the film thickness $\lambda_{P1,3} \gg h$, $\lambda_{\Phi 1,3} \gg h$. One can see from the plot b that the thin region of monoclinic ferroelectric phase exists, in contrast to previous studies [7]. The discrepancy may originate from the difference in material parameters, in particular in $a_{12}^\sigma$.

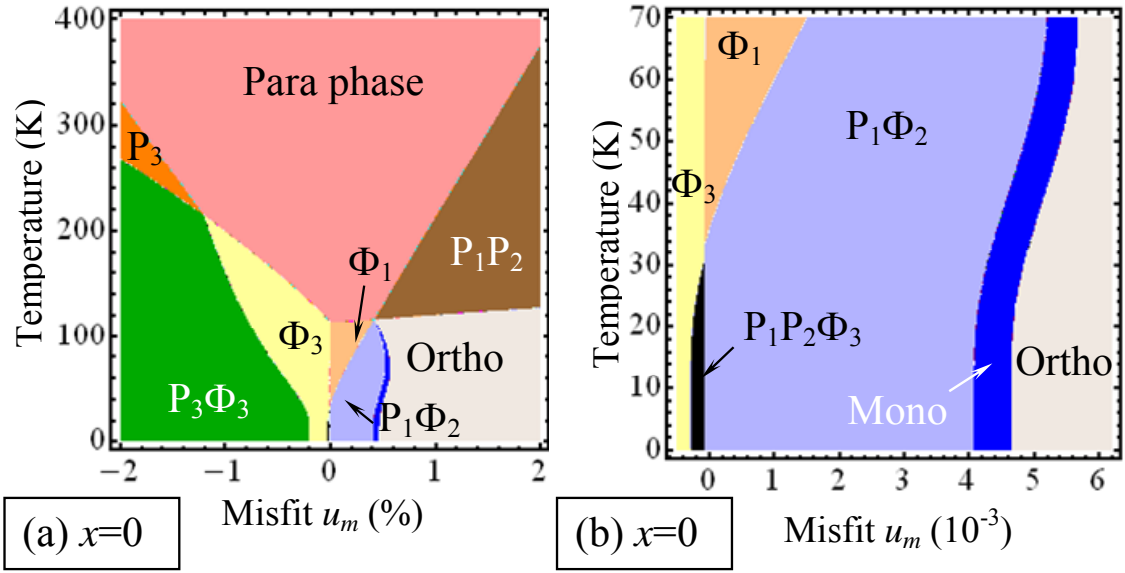

**Figure C1.** Temperature - misfit strain phase diagrams of SrTiO$_3$ thin films for a wide misfit strain range **(a)** and near the small values **(b)**. Abbreviation "mono" stands for the low symmetry monoclinic phase with $P_1 \neq P_2 \neq 0$ and $\Phi_1 \neq \Phi_2 \neq 0$. Abbreviation "ortho" stands for the orthorhombic phase with $P_1 = P_2 \neq 0$ and $\Phi_1 = \Phi_2 \neq 0$.